\documentstyle[psfig]{mn}
\newif\ifAMStwofonts

\ifoldfss
  \ifCUPmtlplainloaded \else
    \NewTextAlphabet{textbfit} {cmbxti10} {}
    \NewTextAlphabet{textbfss} {cmssbx10} {}
    \NewMathAlphabet{mathbfit} {cmbxti10} {} 
    \NewMathAlphabet{mathbfss} {cmssbx10} {} 
  \fi
  \ifAMStwofonts
    \ifCUPmtlplainloaded \else
      \NewSymbolFont{upmath} {eurm10}
      \NewSymbolFont{AMSa} {msam10}
      \NewMathSymbol{\upi}     {0}{upmath}{19}
      \NewMathSymbol{\umu}     {0}{upmath}{16}
      \NewMathSymbol{\upartial}{0}{upmath}{40}
      \NewMathSymbol{\leqslant}{3}{AMSa}{36}
      \NewMathSymbol{\geqslant}{3}{AMSa}{3E}

      \let\leq=\leqslant \let\le=\leqslant
       \let\ge=\geqslant
    \fi
  \fi
\fi 

\ifnfssone
  \newmathalphabet{\mathit}
  \addtoversion{normal}{\mathit}{cmr}{m}{it}
  \addtoversion{bold}{\mathit}{cmr}{bx}{it}
  \newmathalphabet{\mathbfit} 
  \addtoversion{normal}{\mathbfit}{cmr}{bx}{it}
  \addtoversion{bold}{\mathbfit}{cmr}{bx}{it}
  \newmathalphabet{\mathbfss} 
  \addtoversion{normal}{\mathbfss}{cmss}{bx}{n}
  \addtoversion{bold}{\mathbfss}{cmss}{bx}{n}
  \ifAMStwofonts
    \ifCUPmtlplainloaded \else
      \UseAMStwoboldmath
      \makeatletter
      \new@mathgroup\upmath@group
      \define@mathgroup\mv@normal\upmath@group{eur}{m}{n}
      \define@mathgroup\mv@bold\upmath@group{eur}{b}{n}
      \edef\UPM{\hexnumber\upmath@group}
      \new@mathgroup\amsa@group
      \define@mathgroup\mv@normal\amsa@group{msa}{m}{n}
      \define@mathgroup\mv@bold\amsa@group{msa}{m}{n}
      \edef\AMSa{\hexnumber\amsa@group}
      \makeatother
      \mathchardef\upi="0\UPM19
      \mathchardef\umu="0\UPM16
      \mathchardef\upartial="0\UPM40
      \mathchardef\leqslant="3\AMSa36
      \mathchardef\geqslant="3\AMSa3E

      \let\leq=\leqslant \let\le=\leqslant
       \let\ge=\geqslant
    \fi
  \fi
\fi 

\ifnfsstwo
  \DeclareMathAlphabet{\mathbfit}{OT1}{cmr}{bx}{it}
  \SetMathAlphabet\mathbfit{bold}{OT1}{cmr}{bx}{it}
  \DeclareMathAlphabet{\mathbfss}{OT1}{cmss}{bx}{n}
  \SetMathAlphabet\mathbfss{bold}{OT1}{cmss}{bx}{n}
  \ifAMStwofonts
    \ifCUPmtlplainloaded \else
      \DeclareSymbolFont{UPM}{U}{eur}{m}{n}
      \SetSymbolFont{UPM}{bold}{U}{eur}{b}{n}
      \DeclareSymbolFont{AMSa}{U}{msa}{m}{n}
      \DeclareMathSymbol{\upi}{0}{UPM}{"19}
      \DeclareMathSymbol{\umu}{0}{UPM}{"16}
      \DeclareMathSymbol{\upartial}{0}{UPM}{"40}
      \DeclareMathSymbol{\leqslant}{3}{AMSa}{"36}
      \DeclareMathSymbol{\geqslant}{3}{AMSa}{"3E}

      \let\leq=\leqslant \let\le=\leqslant
       \let\ge=\geqslant
    \fi
  \fi
\fi 

\ifCUPmtlplainloaded \else
  \ifAMStwofonts \else 
    \def\upi{\pi}
    \def\umu{\mu}
    \def\upartial{\partial}
  \fi
\fi

\newcommand{\sqdeg}{deg$^2$}
\newcommand{\hMpc}{h$^{-1}$\,Mpc}
\newcommand{\kms}{km\,s$^{-1}$}
\newcommand{\powerten}[1]{$10^{#1}$}
\newcommand{\MB}{M$_{\rm B}$}
\newcommand{\NH}{N$_{\rm H}$}

\title{Detection of the first X-ray selected large AGN group}
\author[Frank Tesch, Dieter Engels]
       {Frank Tesch, Dieter Engels \\
        Hamburger Sternwarte, Gojenbergsweg 112, D-21029 Hamburg, Germany}
\date{Accepted 1988 December 15.
      Received 1988 December 14;
      in original form 1988 October 11}

\pagerange{\pageref{firstpage}--\pageref{lastpage}}
\pubyear{1999}

\begin{document}

\maketitle

\label{firstpage}

\begin{abstract}
We have examined the spatial distribution of 856 AGN detected by the ROSAT
All-Sky Survey (RASS) using a direct search for structures 
with the minimal spanning tree. 
The AGNs were compiled from an area of $\sim$ 7000 deg$^2$, in which optical
identifications of RASS sources were made with the help of the digitized
objective prism plates of the Hamburg Quasar Survey (HQS). Redshifts were 
taken from the literature or from own follow-up observations. The sample
probes the spatial distribution at low redshifts, since the 
redshift distribution peaks at z $\sim$ 0.1. 
The application of the minimal spanning tree led to a 1.8$\sigma$ 
discovery of an 
AGN group with 7 members in a volume \mbox{V $\sim$ 140 $\times$ 75 $\times$ 
75 h$^{-3}$Mpc$^{3}$} in the Pisces constellation. With a mean redshift
z=0.27 this group is only the third discovered group at redshifts z$<$0.5. 
The RASS offers excellent possibilities to study
large scale structure with AGNs at low redshifts, once these redshifts
are determined.

\end{abstract}


\section{Introduction}

Since first indications of cellular structures in the distribution of 
galaxies taken from the 'Second Reference Catalogue' 
were found by Joeveer \& Einasto (1978) more than 20 years ago 
the field of large-scale structure research has been developed 
rapidly. 
Already a few years later, due to the mapping of the nearby universe 
(z $<$ 0.1)
(Gregory \& Thompson 1978, Davis, Huchra \& Latham 1983, De Lapparent, Geller 
\& Huchra 1986),
the last sceptics had to accept the existence of filamentary and cellular 
structures in the spatial distribution of the visible matter.
Nowadays many elements in the network of visible matter like walls,
knots and voids are well examined, 
using galaxies, clusters of galaxies and also superclusters.
Later also Active Galactic Nuclei (AGN) were used to map large scale
structures, beginning with the work by Osmer (1981).
In spite of their lower space density in
comparison to galaxies, significant clustering signals on small scales  
were detected, and also groups of them were discovered. 

\begin{figure*}
\centerline{\psfig{figure=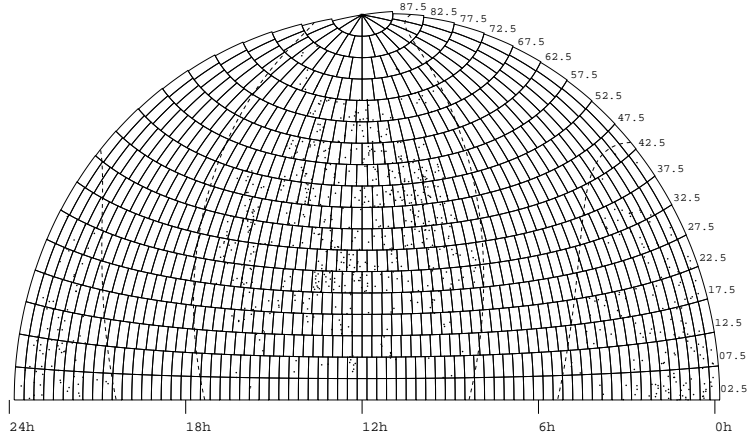,width=10cm}}
\caption[]{\label{om}
Distribution of the RASS-AGNs with known redshifts on the northern hemisphere.
The grid gives the boundaries of HQS fields and the broken lines delineate
latitude $\mid$b$\mid$=20$^\circ$.}
\end{figure*}

The clustering properties of AGNs are of great interest in cosmology
because they allow the study of the evolution of structure over a range
of redshifts $0.3 < {\rm z} < 3$ not accessible with other objects so far.
On small scales ($<$20\,h$^{-1}$Mpc) evidence for AGN clustering 
seems to be well established
(e.g. Shanks et al. 1987, Crampton et al. 1989, 
Iovino, Shaver \& Christiani 1991, Mo \& Fang 1993,
Shanks \& Boyle 1994, Georgantopoulos \& Shanks 1994, Croom \& Shanks 1996),
but its cosmological evolution is discussed controversially
(Iovino \& Shaver 1988, Croom \& Shanks 1996, Kundi\'{c} 1997, 
Stephens et al. 1997, La Franca, Andreani \& Christiani 1998), 
mainly because the clustering properties at low redshifts
are not well determined. 
Sizeable low-redshift AGN samples
require surveys over large areas of the sky and contain presently
less than 200 objects (Boyle \& Mo 1993, Georgantopoulos \& Shanks 1994).
The RASS will provide several thousand new low-redshift AGNs, allowing 
clustering studies in unprecedented details 
as soon as the redshifts will be available. 

In addition to the clustering on small scale, AGN groups with considerably
greater sizes were discovered. 
The first AGN group, detected by Webster (1982) in the CTIO survey, 
contains only 4 members at a redshift of z $\sim$ 0.37. The largest known 
group was 
found in one field of the CFHT grens survey by Crampton, Cowley \& Hartwick 
(1989) comprising 23 AGNs. Clowes and Campusano (1991) searched in the 
direction of 
ESO/SERC field 927 and came upon an elongated group at z $\sim$ 1.3
with 13 members.
Using the minimal spanning tree technique Graham, Clowes \& Campusano
(1995, hereafter GCC)  analyzed several quasar surveys, in which
they could confirm the three already known groups 
and could find another two.
These new groups were detected in the Osmer \& Hewett (1991)
survey at z $\sim$ 1.9 containing 10 members and in the 
Christiani et al.
(1989) and La Franca et al. (1992) survey at z $\sim$ 0.19 with seven members.
The next successful search was done by Komberg, Kravtsov \& Lukash
(1996, hereafter KKT) carrying out
a cluster analysis method known as the friend-of-friend technique 
(Einasto et al. 1994) in the V\'{e}ron-Cetty \& V\'{e}ron Catalogue (1991). 
They found 11 new groups with at least 10 members each and redshifts
greater than 0.6. The group discovered by Crampton et al. was confirmed.
Finally Newman et al. (1997) discovered a group of 13 AGN at z $\approx$
1.51  in the Chile-UK quasar survey.
All in all we are aware of 17 groups of AGN having dimensions 
of 60 - 200 h$^{-1}$Mpc. These sizes are significantly larger than
the typical sizes ($\approx$ 30h$^{-1}$Mpc) of superclusters of galaxies.
Most of them are too far away to allow presently studies of
 their spatial  relation to the associated
distribution of galaxies and clusters of galaxies, 
but it might be possible that they trace superstructures on scales
$\gg 30$ \hMpc\ (KKT).
We present here the discovery of another group, which is the 
first X-ray selected group of AGN.    


\section[]{RASS AGN sample}
The ROSAT All-Sky Survey contains tens of thousands of AGN with z $<$ 0.5
and offers therefore a possibility to investigate the spatial distribution of 
AGN in the nearby universe. Close clusters of quasars could be searched for. 
Until the spring of 1996 identifications of RASS-AGN were carried out in
338 fields ($\sim$ 7000 deg$^{2}$) of the HQS (Bade et al. 1996). Roughly 
3400 AGN  candidates were 
available altogether ($\sim$ 10 AGN/field). For studying the three dimensional 
distribution only objects with known redshifts could be used. For 367 AGN
discovered by the RASS own follow-up spectroscopy was available. 
\mbox{489 RASS} detected AGN from the literature were added, giving a 
total sample of 856 X-ray selected AGN.  
The surface density was $\sim$ 0.12 AGN/deg$^{2}$, which is roughly of the 
same order as the surface densities of optical surveys for low redshifts.
The advantage of the RASS is the much larger sky coverage than the one 
typically obtained
for optical surveys. 
The distribution of the RASS-AGN on the northern sky is shown in Figure 1, the 
redshift distribution in Figure 2.

\begin{figure}
\psfig{figure=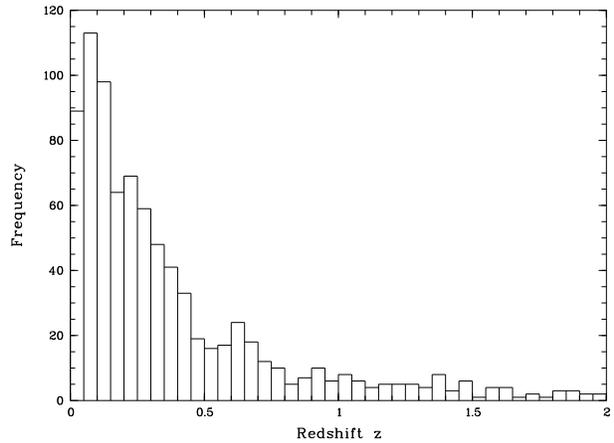,width=8cm}
\caption[]{\label{z}
Reshift distribution of the RASS-AGN sample. 
AGNs with z$>$2.0 (N=21) were omitted.  The highest 
redshift value is 3.62.}
\end{figure}

\section[]{Statistical techniques}
\subsection{Cosmological model, distance determination}
The evaluation of data using cosmological distances requires the
declaration of a cosmological model. We adopted a flat 
universe with $\Omega_{0}$ = 1, $q_{0}$ = 0.5, k = 0 and 
H$_{0}$ = 100\,\kms\,Mpc$^{-1}$.
AGN separations are expressed in 
comoving coordinates. The distance r$_{\rm p}$ 
in comoving coordinates relative to the 
observer, the proper distance, is given by the equation of Mattig (1958):
\begin{displaymath}
r_{p}=\frac{c}{H_{0}q_{0}^{2}(1+z)}\{zq_{0}+(q_{0}-1)[-1+\sqrt{2q_{0}z+1}]\}
\end{displaymath}
The distance R between object 1 and object 2, which are separated by an angle 
$\theta$ ($\theta$=arccos[cos($\Delta \alpha$)cos($\Delta \delta$)]), 
is (Osmer 1981),
\begin{displaymath}
R=\sqrt{D^{2}r_{p_{2}}^{2}+r_{p_{1}}^{2}-2Dr_{p_{1}}r_{p_{2}}cos\theta}
\end{displaymath} 
where 
\begin{displaymath}
D=\sqrt{1-kr_{p_{2}}^{2}}+\frac{r_{p_{2}}}{r_{p_{1}}}cos\theta[1-\sqrt{1-kr_{p_{1}}^{2}}].
\end{displaymath}
For $q_{0}$ = 0.5 and k = 0 we have D = 1, so that the distance R between 
two objects is the cosine rule of Euclidean geometry, i.e., 
\begin{displaymath}
R=\sqrt{r_{p_{1}}^{2}+r_{p_{2}}^{2}-2r_{p_{1}}r_{p_{2}}cos\theta}.
\end{displaymath}

\subsection{Minimal Spanning Tree}
To study the distribution of the RASS-AGN sample we used the minimal
spanning tree (MST) technique which has widespread applications
in many scientific fields. First algorithms were published 
by Kruskal (1956) and Prim (1957). For details on the
historical evolution of the MST technique we refer to Graham and Hell (1985).
GCC introduced a version of this technique into
astronomy which is in particular useful for the purpose of detecting groups 
or clusters of objects in any spatial distribution. They discovered
successfully two new groups of quasars, prompting us to apply this
technique in our case. Crucial steps for this technique are the 
determination of a critical separation distance (Dussert et al. 1987, 
GCC) and of a significance level. 

\begin{figure}
\centerline{\psfig{figure=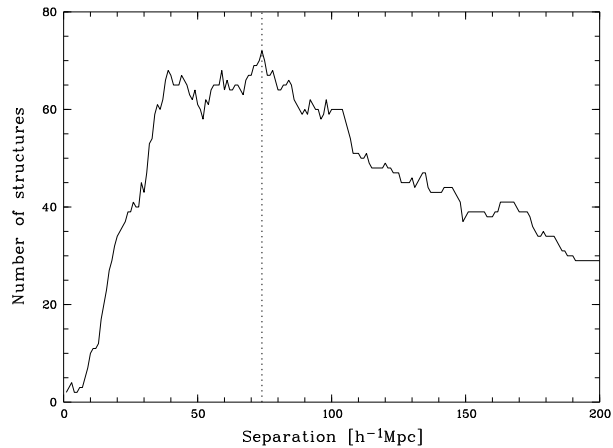,width=8cm}}
\caption[]{\label{ms}
The number of connected structures in the MST
versus the separation distance is shown. 
The value of the separation distance at which the number of structures is
greatest is taken to be the critical separation distance (dashed line at 
74 h$^{-1}$Mpc).}
\end{figure}

The separation distance 
of a given tree is the largest distance allowed to the nearest neighbor.
A choice of this separation distance cuts the tree into connected
structures with at least two members. The critical separation distance
is the optimum value to search for superstructure candidates. 
Figure \ref{ms} shows the number of connected structures as function
of the separation distance. In our case it is a distribution with
an almost flat plateau between 40 and 100 \hMpc. GCC
 used the maximum of this distribution as critical separation
distance, as their distribution was signifcantly peaking at this value.
This is not the case here, but we decided to use the maximum too, as it
is located at  74 \hMpc\ almost in the middle of the plateau. 
The choice of a smaller separation distance as the critical distance would 
result in smaller structures, probably substructures or cores of 
superstructures down to pairs of objects. In the opposite direction 
huge and elongated structures with extensions of a few hundreds of Mpc 
would be considered as structure candidates which are unlikely having a 
physical connection.

To find out, whether a structure is real one has to
determine a significance level. Following GCC, we calculated
the normalized mean m and the 
normalized standard deviation $\sigma$ of the edge-lenghts of the MST  
for each superstructure candidate
and compared them to the values m$_{s}$ and $\sigma_{s}$ obtained from 
randomly 
generated samples. The edge-lengths are here the distances R between the
AGNs. For this purpose {\it 10\,000 random
samples} were generated for each candidate, which have the same number of 
objects and are distributed within a volume of the same size.
The volume is chosen as euclidean rectangular box with
edges defined by the extremes of the $\alpha, \delta,$ z values of the
candidate members.
The significance level is then 
given by P = $\frac{N}{10\,000} \cdot 100$, where N is the number of
 trees with 
m$_{s} \le$ m and $\sigma_{s} \le \sigma$. 

\section[]{Results}
\begin{figure*}
\centerline{\psfig{figure=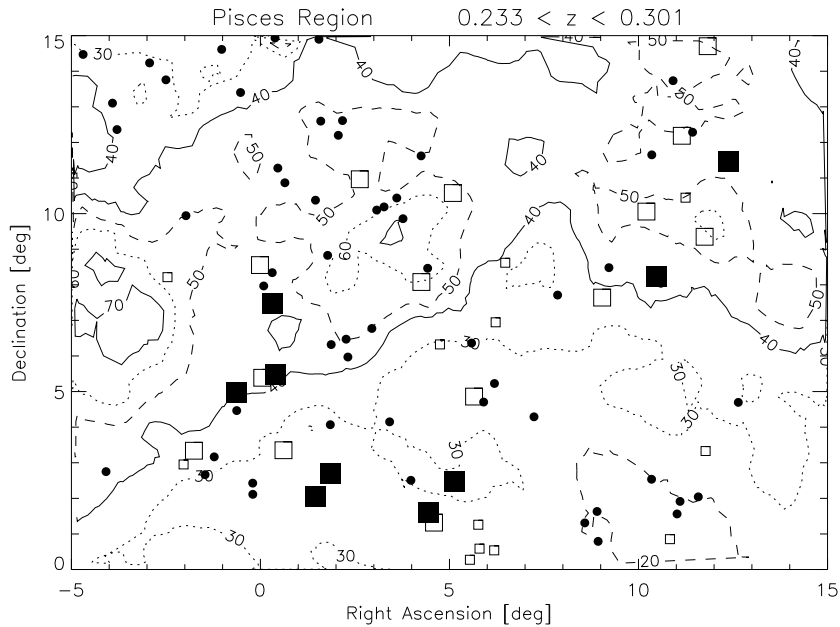,width=12cm}}
\caption[]{\label{pisces} 
Pisces AGN Group (filled squares) at 
$z=0.267\pm0.033$. The two objects to the right are in the same redshift 
range but their relation to the group is unknown. AGN in front of (large 
squares) and behind the group (small squares), and RASS-AGN candidates 
with unknown 
redshift (dots) are coded respectively. North is up and east is left.
The contours represent the hydrogen column densities in units of 
10$^{19}$cm$^{2}$ taken from the Leiden/Dwingeloo Survey.}
\end{figure*}

The application of the MST with a critical separation distance of
74 \hMpc\ on the RASS AGN sample provided seven connected structures
with at least five members each. These seven structures were adopted as
superstructure candidates. The limit of at least five members was adopted,
because the number of connected structures
with less than five members increases very rapidly.
The diameters of the superstructure candidates
were of the order of (100 - 200) h$^{-1}$Mpc, while typical angular
diameters were 5$^{\circ}$ - 10$^{\circ}$.    
The comparison with random generated structures resulted in significance 
levels between 15 and 80 per cent for all but one candidate.
This candidate (Table \ref{p_tab}), 
which we call 'Pisces AGN Group' has a significance 
level of P = 7.5 per cent.
This first X-ray selected AGN group is located at 
$\alpha \approx 0^{h}30^{m}, \delta \approx 5^{0}$, has 7 members at
z = 0.27 $\pm$ 0.03 and covers a volume of $140 \times 75 \times 75$
h$^{-3}$Mpc$^{3}$ (z,$\alpha,\delta$) (Fig. \ref{pisces}). 

The identification of the 'Pisces AGN group' within the MST does not depend
strongly on the choice of the critical distance. 
The MST of the 'Pisces AGN Group' contains six 'edge-lengths' spanning a
 range from 32 to 53 h$^{-1}$Mpc, while several group members have AGN
neighbors with separations of 75--80 h$^{-1}$Mpc. 
Therefore, over a wide range of
choices for the critical distance (53--75 h$^{-1}$Mpc) 
the AGN group would have been recognized,
and only for a choice $<$53 \hMpc\ the group would have fallen apart
into subgroups. However, for 
critical distances $>$75 \hMpc\ the group could have been missed.



\begin{table}
\caption{Pisces AGN Group}
\label{p_tab}
\begin{tabular}{llllll} \hline
RXJ & $\alpha_{2000}$ & $\delta_{2000}$ & z & B & Ref. \\ \hline
23574+0457 & 23 57 28.8 & +04 58 11 & 0.285 & 16.9 & 1\\ 
00012+0728 & 00 01 18.1 & +07 28 26 & 0.270 & 17.9 & 1\\ 
00015+0529 & 00 01 33.4 & +05 29 19 & 0.251 & 16.9 & 1\\ 
00057+0203 & 00 05 47.5 & +02 02 59 & 0.234 & 16.5 & 2\\
00074+0241 & 00 07 27.1 & +02 41 12 & 0.300 & 17.6 & 2\\ 
00177+0136 & 00 17 45.8 & +01 36 20 & 0.236 & 18.2 & 2\\ 
00205+0228 & 00 20 33.2 & +02 28 52 & 0.256 & 17.9 & 2\\ \hline
\end{tabular}\medskip\\
References : 1) own follow-up spectroscopy, 2) Hewett et al. 1995\\
\end{table}

Shape and size of the group as well as the ability to detect further
members in soft X-rays may depend on variations in sensitivity of the RASS 
across the region. While the exposure times are fairly homogeneous
in the area shown in Fig \ref{pisces}, variations of the hydrogen
column density \NH\ by a factor of 5.6 are present. To
visualize it, we overplotted in Fig \ref{pisces} the \NH\ distribution
as taken from the Leiden/Dwingeloo Survey (Hartmann \& Burton 1997).
There is no obvious clustering of RASS AGN (candidates) in low \NH\ regions
nor do they avoid systematically regions of higher column density.
Taking the \NH\ variations into account we find that the RASS
sensitivity varies at most by {\it 18 per cent} throughout the region. 
As the peak
values of \NH\ are restricted to rather small areas, we conclude that
losses introduced by variations of the RASS sensitivity are negligible. 
Note that several dozen AGN candidates are already known in the Pisces
region, which we could not take into account because of their unknown
redshifts.

The 'Pisces AGN Group' is only the third large AGN group discovered
at z$<$0.5, and the first found by an X-ray survey. The group contains
AGNs with luminosities -22.5 $\le$\MB$\le$ -24.4 around the \MB=-23.0 limit
at which AGNs are separated into Seyfert galaxies and quasars. The other
two low-redshift large agglomerations are the group of four AGNs at z=0.37 
(-22.8 $\le$\MB$\le$ -24.4) analyzed by Webster \cite{Webster82}, and
the group of seven Seyfert galaxies (-20.4 $\le$\MB$\le$ -22.3) at
z=0.19 discovered by GCC. Their sizes
of $95\times30\times15$ h$^{-3}$ Mpc$^3$ and 
$60\times30\times10$ h$^{-3}$ Mpc$^3$ are considerably
smaller than the size of the 'Pisces AGN Group' resulting in higher
number densities for the former. The correponding numbers are 
9$\cdot$\,\powerten{-5}, 
39$\cdot$\,\powerten{-5}, and
1$\cdot$\,\powerten{-5} h$^3$ Mpc$^{-3}$. With a level of 7.5 per cent the 
significance
of the 'Pisces AGN Group' is still lower than for the other groups
(P $<$ 1 per cent) making a verification by completing the
 redshift determinations
of the RASS-AGN candidates in this region (cf. Fig. \ref{pisces}) desirable.
We note that at a distance of $\approx$10$^{\circ}$ ($\approx$ 120 \hMpc)
of the group two AGNs are known within the groups redshift range.
If further AGNs are found in the region in between, the 'Pisces
AGN Group' might be part of an even larger structure. Perhaps it
is a node in a filamentary AGN superstructure extending north-west from the
group (cf. Fig. \ref{pisces}). 

\section[]{Discussion}
The discovery of the Pisces Quasar Cluster in our rather inhomogeneously
distributed sample encourages to use the RASS for further studies of
the large scale distribution of low-redshift AGN. The full information
however will be available only after a complete determination of
redshifts of all the RASS-AGN candidates, at least on a part of the sky.
{\it From the
identification} of RASS sources on Hamburg Schmidt Plates follows,
that about {\it 50 per cent} of the sources 
with $\mid b \mid > 30^{\circ}$ are AGNs
(Bade et al. 1998), so that
a surface density of $\approx$ 0.7 AGN/deg$^{2}$ can be reached. 
Based on the distribution of redshifts of
RASS-AGNs from Bade et al. \cite{Bade95} (see also Fig. \ref{z}), 
the expected surface density for $z \leq 0.5$ 
is still 0.5 AGN/deg$^2$. 
Such a surface density has not been reached by any 
other non-optical survey in this redshift range so far. 
Only optical surveys may reach similar surface
densities, but their areas searched are significantly smaller and their
efficiency {\it to find AGN are much lower}.
For example, the largest of them, the 'Large Bright Quasar Survey' (LBQS), 
discovered 1055
quasars on an effective area of 454 deg$^{2}$ (Hewett et al. 1995).
Among them are 
165 with low redshift ($ 0.2 \le z \leq 0.5$), giving a
surface density of 0.36 AGN/deg$^{2}$. 

Today we consider the RASS as first choice, if searches for structures on
large scales at low redshifts are attempted.
 At z = 0.3 a linear size of 
120 h$^{-1}$Mpc for example corresponds to an angle on the sky of $\sim$
12$^{\circ}$ and at $z=0.1$ to 27$^{\circ}$. Optical surveys the
size two or three times the LBQS would be required to be competitive.
New AGN groups at low redshifts also provide a better chance
to study the underlying matter distribution than at high redshifts,
because the discovery of associated clusters of galaxies or even 
individual galaxies
will be in reach of present day telescopes.

In any case, the nature of AGN groups is not yet well understood.
When Webster \cite{Webster82} found a very low probability that the z=0.37
group of quasars is a chance event, this group with a size of
$\approx$100 \hMpc\ was the largest known structure in the universe.
Meanwhile further quasar groups have been found and evidence from other
observations point out that large scale structure on scales of
$\approx$100 \hMpc\ might be common (Eisenstein et al. \cite{Eisenstein98}
and references therein). Thus, the relation of quasar groups to
the overall matter distribution on large scales is of eminent interest.
Due to the large distances between group
members and therefore because of the low volume density, 
it must be assumed that groups of AGN are 
not held together by gravitational forces.
Presupposed that these
structures are real, such groups could trace the  structure
of the universe at rather large scales, 
and are possibly embedded in the spatial distribution
of galaxies and clusters of galaxies. Up to now their role in the  
network of galaxies is absolutely unknown.

KKT suggest that the Large Quasar Groups at higher z belong to
concentrations of young galaxy clusters and evolve into the known
superclusters of today. Quasar groups would point to sites of
enhanced matter density. They derive a local spatial number density
of superclusters of n$_{\rm{SCL}} \approx$ 1.4$\cdot$\,\powerten{-7} h$^3$ Mpc$^{-3}$
and predict about the same number of Large Quasar Groups at higher 
redshifts as they actually found in a number of homogeneous
quasar surveys. 

Their evolutionary scenario predicts several dozen new quasar
groups in the volume accessible now with the help of the RASS.
With an area of $\Omega\approx$7000 \sqdeg\ in which RASS identifications
were available, and restricting to a redshift range 
0.05$\le$z$\le$0.3, in which the RASS is most sensitive, N$\approx$40
quasar groups with more than ten members are predicted. Thus 
$\ge$400 AGN belonging to quasar groups should be contained in this
volume, making up 12 per cent of all RASS-AGN candidates in this area.
As only part of the group members will actually be detected by the RASS, the
finding of only one significant group (but having 7 members only) in our 
sample might not be at odds with the prediction. One reason for this
disagreement could be that KKT did their estimation with objects
taken from the V\'{e}ron-Cetty \& V\'{e}ron Catalogue (1991) in which
mostly optically selected AGNs are contained. X-ray selected AGNs, however, 
may
have different clustering properties as optically selected AGNs. 
Carrera et al. (1998) presented a first clustering analysis of X-ray
selected AGNs taken from the RIXOS Survey. They obtained limits
on the AGN clustering estimating the correlation length $r_{0}$, and their
values showed consistency with the clustering of galaxies but not with 
clustering of 
optically selected AGNs in the way that X-ray selected AGNs are less
clustered. 
Furthermore, follow-up spectroscopy of RASS-AGN candidates is rather 
incomplete. {\it For only 
about one fourth} of the AGN candidates 
the redshifts are known so far. These lacking redshifts will probably not 
explain the disagreement with the KKT hypothesis completely but at least 
some AGN groups might be found additionally.         

Because of their high luminosity, quasars are excellent tools to study
large scale structure beyond the limits of the deepest wide-angle galaxy 
redshift
surveys. If quasar activity in galaxies is of too short duration
or occur only in a small fraction of galaxies, {\it their frequency might be}
too low for some of them to be present at a given time in a particular
structure to form a detectable group. This bias factor might prevent
that most of the underlying matter distribution is traced efficiently by
AGN. To determine this bias factor the number density of AGN
groups has to be known. An upper limit might be given by the number density
of the superstructures in the local universe (KKT). A direct determination
would be possible on the base of complete RASS-AGN redshift surveys,
possibly starting with selected areas. For the beginning we started
such a survey on $\approx$ 1600 deg$^{2}$ in three areas on the northern
sky (Engels et al. 1998; see also: www.hs.uni-hamburg.de/rosac.html).

\section[]{Conclusions}
We discovered the first RASS selected group of AGNs
(Pisces AGN Group), showing that the RASS indeed can be used to study large 
scale structure formation at z $<$ 0.5 with AGNs. More groups should have been
discovered, if such groups were associated with structures, 
which develop into the superclusters of galaxies in the local universe.

One can probably search efficiently for superstructures 
(groups, voids, filaments) in the distribution
of RASS AGNs by simply increasing their surface density. This
requires redshift determinations for the RASS AGN candidates,
already pre-identified on Hamburg Schmidt plates.
Follow-up spectroscopy of these candidates will be an efficient way to 
create homogeneous selected samples of low-redshift
AGNs covering a large area on the sky.  With such a sample it should be
possible to determine the volume density of AGN groups.
This will clarify, whether the AGN groups are isolated density peaks in the
spatial distribution of AGNs or whether they belong to rather regular
AGN superstructures. The Sloan Digital Sky Survey will reach the same 
range of redshifts with clusters of galaxies. A comparison of the 
distributions of these
clusters of galaxies with the RASS AGNs should give hints whether both classes
of objects trace the same large scale structures in the universe. 

\section*{Acknowledgments}
The Hamburg/RASS identification program is supported by the Deutsche
Forschungsgemeinschaft (DFG) grant Re\,352/22 and by the
BMBF grant DARA 50 OR 96016. We acknowledge the support of this
work by DFG grant En\,176/13.

\appendix

\bsp

\label{lastpage}


\begin{thebibliography}{99}

\bibitem[1995]{Bade95}
Bade N., Fink H.H., Engels D., Voges W., Hagen H.-J., Wisotzki L., Reimers D.,
 1995, A\&AS 110, 469
\bibitem[1996]{Bade96}
Bade N., Engels D., Voges W., Reimers D., 1996, MPE-Report 263, 647
\bibitem[1998]{Bade98}
Bade N., Engels D., Voges W., Beckmann V., Boller T., Cordis L., Dahlem M., 
Englhauser J., Molthagen K., Studt J., Reimers D., 1998, A\&AS 127, 145
\bibitem[1993]{Boyle93}
Boyle B. J., Mo H. J. 1993, MNRAS, 260, 925
\bibitem[1998]{Carrera98}
Carrera F.J., Barcons X., Fabian A.C., Hasinger G., Mason K.O., McMahon P.G., 
Mittaz J.P.D., Page M.J., 1998, MNRAS 299, 229
\bibitem[1989]{Christiani89}
Christiani S., Barbieri C., Iovino A., La Franca F., Nota A., 1989, 
A\&AS 77, 161  
\bibitem[1991]{Clowes91}
Clowes R. G., Campusano L. E., 1991, MNRAS 249, 218
\bibitem[1989]{Crampton89}
Crampton D., Cowley A.P., Hartwick F.D.A., 1989, ApJ 345, 59  
\bibitem[1996]{Croom96}
Croom S.M., Shanks T., 1996, MNRAS 281, 893
\bibitem[1983]{Davis1983}
Davis M., Huchra J., Latham D., 1983, IAU Symposium 104, in {\em The Early
Evolution of the Universe}
\bibitem[1986]{DeLapparent86}
De Lapparent V., Geller M., Huchra J., 1986, ApJ 302, L1
\bibitem[1987]{Dussert87}
Dussert C., Rasigni M., Palmari J., Rasigni G., Llebaria A., Marty F. 1987,
J. theor. Biol., 125, 317
\bibitem[1994]{Einasto94}
Einasto M., Einasto J., Tago E., Dalton G. B., Andernach H., 1994, 
MNRAS 269, 301 
\bibitem[1998]{Eisenstein98}
Eisenstein D.J., Hu W., Silk J., Szalay A.S., 1998, ApJ 494, L1 
\bibitem[1998]{Engels98}
Engels D., Tesch F., Ledoux C., Wei J., Ugryumov A., Valls-Gabaud D., Hu J., Voges W., 1998, MPE-Report in press (ph/9811182)
\bibitem[1994]{Georgantopoulos94}
Georgantopoulos I., Shanks T., 1994, MNRAS 271, 773 
\bibitem[1985]{Graham85}
Graham R. L., Hell P., 1985, Annals of the History of Computing, 
Volume 7, Number 1
\bibitem[1995]{Graham95}
Graham M. J., Clowes R. G., Campusano L. E., 1995, MNRAS, 275, 790
\bibitem[1978]{Gregory78}
Gregory S.A., Thompson L.A., 1978, ApJ 222, 784
\bibitem[1995]{Hewett95}
Hartmann D., Burton W.B., 1997, Cambridge University Press 1997
\bibitem[1997]{Hartmann97}
Hewett P.C., Foltz C.B., Chaffee F.H., 1995, AJ 109, 1498
\bibitem[1988]{Iovino88}
Iovino A., Shaver P., 1988, ApJ 330, L13
\bibitem[1991]{Iovino91}
Iovino A., Shaver P., Christiani S., 1991, in {\em The Space Distribution of
Quasars}, ASP Conference Series, vol. 21, p. 202
\bibitem[1978]{Joeveer78}
Joeveer M., Einasto J., 1978, IAU Symposium 79, in {\em The Large-Scale 
Structure of the Universe}, p. 241
\bibitem[1996]{Komberg96}
Komberg B.V., Kravtsov A.V., Lukash V.N., 1996, MNRAS 282, 713 
\bibitem[1956]{Kruskal56}
Kruskal J. B., 1956, Proc. Am. Math. Soc. 7, 48
\bibitem[1997]{Kundic97}
Kundi\'{c} T., 1997, ApJ 482, 631
\bibitem[1992]{LaFranca92}
La Franca F., Christiani S., Barbieri C., 1992, AJ 103, 1062
\bibitem[1998]{LaFranca98}
La Franca F., Andreani P., Christiani S., 1998, ApJ 497, 529
\bibitem[1958]{Mattig58}
Mattig W., 1958, Astron. Nachr. 284, 109
\bibitem[1993]{Mo93}
Mo H.J., Fang L.Z., 1993, ApJ 410, 493
\bibitem[1997]{Newman97}
Newman P. R., Clowes R. G., Campusano L. E., Graham M. J., 1997, ph/9710281
\bibitem[1981]{Osmer81}
Osmer P.S., 1981, ApJ 247, 762
\bibitem[1991]{Osmer91}
Osmer P. S., Hewett P. C., 1991, ApJS 75, 273
\bibitem[1957]{Prim57}
Prim R. C., 1957, Bell Sys. Tech. J. 36, 1389
\bibitem[1987]{Shanks87}
Shanks T., Fong R., Boyle B.J., Peterson B.A., 1987, MNRAS 227, 739
\bibitem[1994]{Shanks94}
Shanks T., Boyle B.J., 1994, MNRAS 271, 753
\bibitem[1997]{Stephens97}
Stephens A.W., Schneider D.P., Schmidt M., Gunn J.E., Weinberg D.H., 1997,
AJ 114, 41
\bibitem[1991]{Veron91}
V\'{e}ron-Cetty M.-P., V\'{e}ron P., 1991, A Catalogue of Quasars and Active
Nuclei, ESO Sci. Rep. No. 10.
\bibitem[1982]{Webster82}
Webster A., 1982, MNRAS 199, 683 



\end{thebibliography}
\end{document}